\documentclass[prd,twocolumn,showpacs,aps,epsbox,groupedaddress,eqsecnum]{revtex4}
\usepackage{amsmath,amssymb}
\usepackage[dvipdf]{graphicx}
\usepackage{natbib}
\usepackage{color}
\begin{document}
\draft
\title{Wormhole Shadows in Rotating Dust}
\author{Takayuki Ohgami}
\email{v501wa@yamaguchi-u.ac.jp}
\author{Nobuyuki Sakai}
\email{nsakai@yamaguchi-u.ac.jp}
\affiliation{Graduate School of Science and Engineering, Yamaguchi University, Yamaguchi 753-8512, Japan}

\begin{abstract}
%%%%%%%%%%%%%%%%%%%%%%%%%%%%%%%%
As an extension of our previous work, which investigated the shadows of the Ellis wormhole surrounded by nonrotating dust,
in this paper we study wormhole shadows in rotating dust flow.
First, we derive steady-state solutions of slowly
rotating dust surrounding the wormhole by solving relativistic Euler equations.
Solving null geodesic equations and radiation transfer equations, we investigate the images of the wormhole surrounded by dust for the above steady-state solutions.
Because the Ellis wormhole spacetime possesses unstable circular orbits of photons, a bright ring appears in the image, just as in Schwarzschild spacetime.
The bright ring looks distorted due to rotation.
Aside from the bright ring, there appear weakly luminous complex patterns by the emission from the other side of the throat.
These structure could be detected by high-resolution very-long-baseline-interferometry observations in the near future.
\end{abstract}
%%%%%%%%%%%%%%%%%%%%%%%%%%%%%%%%
\pacs{04.40.-b, 97.60.Lf}
\maketitle

\section{Introduction}
A wormhole is a tunnel-like structure which connects two distant or disconnected regions.
A spacetime with nontrivial topology such as a wormhole is approved by general relativity and other extended gravitational theories.
The Einstein-Rosen bridge, which is considered as a first wormhole solution of Einstein equations, was discovered by Einstein and Rosen.
Because this wormhole is not traversable, it was regarded as nothing but a mathematical product \cite{ER}.
Several decades later, Ellis \cite{Ellis} obtained a new wormhole solution: a spherically symmetrical solution of Einstein equations with a ghost massless scalar field.
Morris and Thorne \cite{005} showed that the Ellis wormhole is one of the traversable wormholes.
These wormholes have neither singularity nor horizon; their tidal force is so weak that people can withstand. 
If such wormholes exist, they could become a fascinating tool for voyaging to far galaxies or engaging in time travel by passing through that.

The stability of traversable wormholes such as Ellis wormholes have been studied by several researchers.
Shinkai and Hayward \cite{SH} showed the instability
of Ellis wormhole by numerical simulations.
Gonz\'alez \textit{et al.} \cite{GGS} considered the more general wormholes with the ghost scaler field are also unstable.
These researches indicated that Ellis wormholes and other traversable wormhole with a ghost scaler field are practically nonexistent.
However, Das and Kar \cite{DK} pointed out that another matter could contribute to supporting the Ellis geometry.
Furthermore, under the modified gravitational theories, matter such as a ghost scaler field, which makes wormhole spacetimes unstable, may not be required.
Therefore, traversable wormholes are still a viable subject not only in theoretical physics, but also in observational astrophysics.

A possible method for probing wormholes is based on gravitational lensing effects.
Basic properties of their gravitational lensing effects were investigated theoretically in Ref.\cite{basic}.
Since Cramer \textit{et al.} \cite{001} pointed out anomalous features of the light curve of a distant star lensed by a wormhole, observational research to find wormholes by using the microlensing effect has proceeded \cite{002}.
In addition to the light curve, the lensed images \cite{image} and the lensed spectra \cite{spectra} of Ellis wormholes have also been discussed as observable quantities.

In a case of probe for blackholes by electromagnetic observations, another method is the usage of shadows, which are the images of optical or radio sources around a blackhole.
Blackhole shadows were originally discussed by Bardeen \cite{Bardeen} and have recently attracted much attention \cite{Takahashi}.
This phenomenon has been researched not only theoretically but also observationally by very-long-baseline-interferometry (VLBI) for probing blackholes \cite{VLBI}.
Therefore, we expect shadows as many properties of probing Ellis wormholes by use of VLBI observations.
Nedkova, Tinchev and Yazadjiev researched shadows caused by a rotating wormhole \cite{The Shadow of a Rotating Traversable Wormhole}.
They calculated the outline of wormhole in the widely distributed light source.

It is known that photon orbit on the Ellis wormhole spacetime has unstable circular orbits.
The existence of this orbit is important for optically observations and the brightly ring appears in the optical image of the wormhole surrounded by the optically thin light source: we discuss in Sec. III.
We have considered a case that dust has radial momentum only, and computed the optical image of the wormhole surrounded by optically thin dust \cite{spherical steady-state solutions}.
In this paper, we assume that dust has not only radial momentum but also angular momentum.
This assumption is more natural state than the former case. 

This paper is organized as follows.
In Sec. II, we introduce the Ellis wormhole and discuss its spacetime structure.
In Sec. III, we derive null geodesic equations and discuss photon trajectories around the Ellis wormhole.
In Sec. IV, to set up dust models used in our shadow analysis, we derive steady-state solutions of dust surrounding the wormhole by solving relativistic Euler equations.
In Sec. V, we investigate---by solving the radiative transfer equation along the null geodesic---the images of wormhole surrounded by dust for the models obtained in Sec. IV.
Section VI is devoted to concluding remarks.
%%%%%%%%%%%%%%%%%%%%%%%%%%%%%%%%
%
%
%
%
%
%%%%%%%%%%%%%%%%%%%%%%%%%%%%%%%%
\section{Ellis wormhole}
\begin{figure}[h]
	\centering
	\includegraphics[width=.4\hsize]{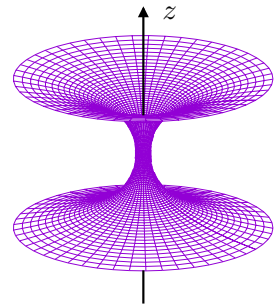}
	\caption{Diagram of two-dimensional Ellis wormhole curve surface embedded in three-dimensional Euclidean space: plot of $z$ vs. $r^*$ in Eq.(\ref{eq:relation between z and r*}) with the $\varphi$ direction. It shows that two various spaces are connected by the Ellis wormhole like a tunnel.}
	\label{fig:2D wormhole image}
\end{figure}

The Ellis wormhole is the spacetime structure discussed by the line element
\begin{align}
	ds^2=-dt^2+dr^2+(r^2+a^2)(d\theta^2+\sin^2\theta\,d\varphi^2),
	\label{eq:line element of Ellis wormhole spacetime}
\end{align}
where $a$ is the throat radius of wormhole and we adopt the unit system of $c=1$.
People can pass through this wormhole alive because it has neither horizon nor singularity.
We draw the embedded diagram of the two-dimensional wormhole surface in the three-dimensional Euclidean space to understand visualy this spacetime structure.
Introducing a new radial coordinate,
\begin{equation}
r^*\equiv\sqrt{r^2+a^2},
\end{equation}
we express the two-dimensional wormhole surface of $t,\,\theta=\text{const.}$ by the line element
\begin{align}
	{ds_{\text{WH}}}^2=\frac{1}{1-a^2/{r^*}^2}{dr^*}^2+{r^*}^2d\varphi^2.
	\label{eq:line element of Ellis wormhole curved surface}
\end{align}
And then, the line element of three-dimensional Euclidean space described by cylindrical coordinates is given by
\begin{align}
	{ds_{\text{ES}}}^2=dz^2+{dr^*}^2+{r^*}^2d\varphi^2.
	\label{eq:line element of Euclidean space by cylindrical coordinates}
\end{align}
Assuming these line elements are equal ${ds_{\text{WH}}}={ds_{\text{ES}}}$, we obtain the relation
\begin{align}
	z=\pm a\cdot \text{arccosh}\frac{r^*}{a}.
	\label{eq:relation between z and r*}
\end{align}
Figure \ref{fig:2D wormhole image} is a graph of Eq.(\ref{eq:relation between z and r*}) with the $\varphi$ direction, which indicates the visual image of the spacetime structure of the Ellis wormhole.
Two various spaces are connected by the Ellis wormhole like a tunnel.
If the wormhole exists in our three-dimensional space, we observe the throat as ball shaped like structure.

\section{Photon trajectories}
In preparation to investigate shadow phenomena, we review basic properties  of the photon trajectories, which was discussed in Ref.\cite{orbit,spherical steady-state solutions}.

\subsection{Null geodesic equations and effective potential}
\begin{figure}[h]
	\centering
	\includegraphics[width=1.0\hsize]{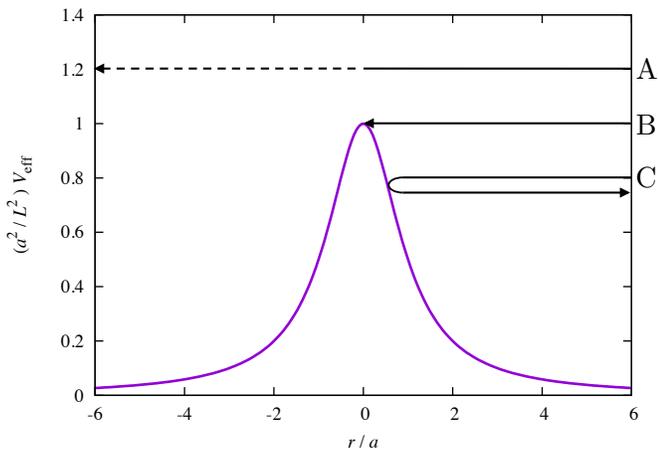}
	\caption{Effective potential of photons in the Ellis wormhole spacetime. The maximum point at $r=0$, which indicates the throat of the wormhole, corresponds to the unstable circular orbits. The region for by $r<0$ is the other side of the spacetime and we use a dashed line for the trajectory in this region.}
	\label{fig:effective potential of WH}
\end{figure}

Null geodesic equations are generally described as
\begin{align}
	\frac{dk^{\mu}}{d\lambda}+\Gamma^{\mu}_{\nu\sigma}k^{\nu}k^{\sigma}=0,\quad 
	\text{with}\quad
	k_{\mu}k^{\mu}=0,
	\label{eq:general null geodesic equations}
\end{align}
where $\lambda,\,k^{\mu}\equiv dx^{\mu}/d\lambda$ and $\Gamma^{\mu}_{\nu\sigma}$ are the affine parameter, the null vector and the Christoffel symbol, respectively.
The null geodesics in the $\theta=\pi/2$ plane are given by
\begin{align}
	&\frac{dk^t}{d\lambda}=0,\quad \frac{d}{d\lambda}\left\{ (r^2+a^2)k^{\varphi} \right\}=0, 
	\label{eq:null geodesic eq. in theta=pi/2 1}\\
	&\frac{dk^r}{d\lambda}-r{k^{\varphi}}^2=0, 
	\label{eq:null geodesic eq. in theta=pi/2 2}\\
	&-{k^t}^2+{k^r}^2+(r^2+a^2){k^{\varphi}}^2=0.
	\label{eq:null geodesic eq. in theta=pi/2 3}
\end{align}
We need not solve Eq.(\ref{eq:null geodesic eq. in theta=pi/2 2}) because it can be derived from Eq.(\ref{eq:null geodesic eq. in theta=pi/2 1}) and (\ref{eq:null geodesic eq. in theta=pi/2 3}).
Integrating Eq.(\ref{eq:null geodesic eq. in theta=pi/2 1}) with respect to the affine parameter, we obtain conserved quantities $E$ and $L$ as
\begin{align}
	E=k^t,\quad L=(r^2+a^2)k^{\varphi}.
	\label{eq:conserved quantities}
\end{align}
Plugging these quantities into Eq.(\ref{eq:null geodesic eq. in theta=pi/2 3}), we derive
the equation corresponding to the energy conservation as
\begin{align}
	{k^r}^2+V_{\text{eff}}(r)=E^2,\quad
	V_{\text{eff}}(r)\equiv\frac{L^2}{r^2+a^2}.
	\label{eq:energy conserving eq.}
\end{align}
Figure \ref{fig:effective potential of WH} shows the graph of the effective potential $V_{\text{eff}}$.
This graph is useful for discussing photon trajectories.

\subsection{Photon trajectories}
The trajectories are classified into three types A, B, and C, as shown in Fig.\ref{fig:effective potential of WH}.
Type A trajectories do not hit the wall of the potential and go into the region of $r<0$ from that of $r>0$;
this means that the photons pass through the wormhole into the other side.
Type B trajectories approach the local maximum of the potential;
this means that the the photons rotate around the throat infinite times in approaching the throat.
Considering time reversal, we can interpret that any photon which rotates around $r=0$ is unstable and eventually goes into $r\rightarrow\infty$ or $r\rightarrow-\infty$.
Type C trajectories bounce from the wall of potential and goes to infinity;
in this case light rays are refracted by gravity.

\begin{figure}[h]
	\centering
	\includegraphics[width=0.75\hsize]{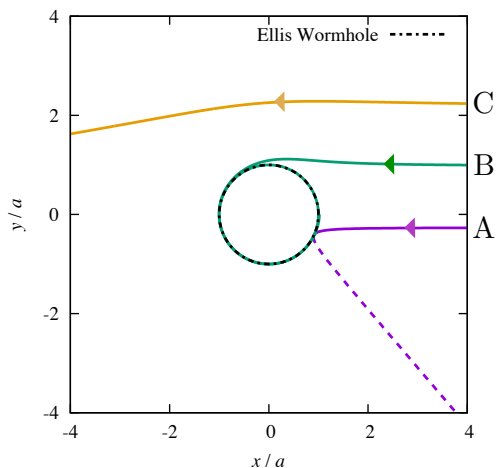}
	\caption{Photon trajectories around the Ellis wormhole. The coordinates $(x,\,y)$ are defined as (\ref{eq:rectangle coordinates}). The dashed circle denotes the throat of the wormhole. All trajectories end up at the observer at $r=300a,\, \varphi =0$. Three labels A, B, and C correspond to those in Fig.\ref{fig:effective potential of WH}. The dashed line of the label A represents the trajectory on the other side region of the wormhole $(r<0)$.}
	\label{fig:photon trajectories around Ellis WH}
\end{figure}

We show the photon trajectories around the Ellis wormhole in Fig.\ref{fig:photon trajectories around Ellis WH}.
Here we define the rectangle coordinates on the $\theta=\pi/2$ plane as
\begin{align}
	x=r^*\cos\varphi,\quad
	y=r^*\sin\varphi,\quad
	r^*=\sqrt{r^2+a^2}
	\label{eq:rectangle coordinates}
\end{align}
These orbits start at the point $r=300a$ and $\varphi=0$.
The three labels A, B and C correspond to those in Fig.\ref{fig:effective potential of WH}.
Furthermore, the dashed line labeled A represents the photon trajectory in other side.

\section{dynamics of interstellar medium}

In this section, we examine the general relativistic motion of interstellar medium surrounds the Ellis wormhole.
Here we assume that the interstellar medium is the perfect fluid and its self-gravity is negligible.
The equations of motion of the perfect fluid in Schwarzschild spacetime and their solutions were presented in a textbook by Shapiro and Teukolsky \cite{BLACK HOLES WHITE DWARFS AND NEUTRON STARS THE PHYSICS OF COMPACT OBJECTS}.
The dynamics of nonrotating fluid around the Ellis wormhole was discussed in Ref.\cite{spherical steady-state solutions}.

\subsection{Thermodynamics}
Suppose a local Lorentz frame comoving with fluid particles.
Let $n,\,\rho$ and $P$ be, respectively, the number density, the total energy density, and the pressure of the fluid particles, which are measured in the reference frame.
Then the first law of thermodynamics is written as
\begin{align}
	dQ=d\left( \frac{\rho}{n} \right)+Pd\left( \frac1n \right),
	\label{eq:first law of thermodynamics}
\end{align}
where $dQ$ is the heat gained per particle.
$\rho/n$ and $1/n$ represent, respectively, the energy and the volume per particle.
We use the temperature $T$ and the entropy per particle $s$ and $dQ=Tds=0$ is true on the process of quasistatic (i.e., in thermal equilibrium at all times) and adiabatic.
Thus, we obtain
\begin{align}
	0=d\left( \frac{\rho}{n} \right)+Pd\left( \frac1n \right),
	\label{eq:first law of thermodynamics 2}
\end{align}
We rewrite this equation in the following, simpler form:
\begin{align}
	\frac{d\rho}{dn}=\frac{\rho+P}{n}.
	\label{eq:first law of thermodynamics 3}
\end{align}

\subsection{Fluid dynamics}
Under the assumption that the particle number is conserved, we obtain the general relativistic continuity equation,
\begin{align}
	(nu^{\mu})_{;\mu}=0,
	\label{eq:general relativistic continuity equation}
\end{align}
where $u^{\mu}$ is the four-velocity of the particle fluid and the semicolon denotes a covariant derivative: $A^{\alpha}_{;\beta}=\partial_{\beta}A^{\alpha}+\Gamma^{\alpha}_{\beta \gamma}A^{\gamma}$.
Moreover, the energy-momentum tensor is written as
\begin{align}
	T^{\mu\nu}=(\rho+P)u^{\mu}u^{\nu}+Pg^{\mu\nu},
	\label{eq:energy momentum tensor for perfect fluid}
\end{align}
because we assume that the particle fluid is a perfect fluid.
Applying the energy-momentum conservation,
\begin{align}
	T^{\nu}_{\mu;\nu}=0,
	\label{eq:conservation law of energy-momentum}
\end{align}
to Eq.(\ref{eq:first law of thermodynamics 3}) with Eq.(\ref{eq:energy momentum tensor for perfect fluid}), we obtain the general relativistic Euler equation,
\begin{align}
	(\rho+P)u_{\mu;\nu}u^{\nu}=-P_{,\mu}-u_{\mu}P_{,\nu}u^{\nu}.
	\label{eq:general relativistic Euler eq.}
\end{align}
We have two basic equations (\ref{eq:general relativistic continuity equation}) and (\ref{eq:general relativistic Euler eq.}), for the general perfect fluid.

\subsection{Slowly rotating dust solutions}
The spherically symmetrical and steady-state solutions were derived in Ref.\cite{spherical steady-state solutions}.
In this paper we extend our analysis to nonspherical (axially symmetric) cases.
We suppose that the four-velocity and number density takes the forms of
\begin{align}
	&u^{\mu}=\left(u^t(r,\,\theta),\,u^r(r,\,\theta),\,u^{\theta}(r,\,\theta),\,u^{\varphi}(r,\,\theta)\right),
	\label{eq:four velosity}\\
	&n=n(r,\,\theta),
	\label{eq:number density}
\end{align}
We assume that the interstellar medium is only dust,
\begin{align}
	\rho=mn\quad\text{and}\quad P=0,
	\label{eq:dust approximation}
\end{align}
where $m$ is the mass of the fluid particle.
Then, we can rewrite Eq.(\ref{eq:general relativistic continuity equation}) and (\ref{eq:general relativistic Euler eq.}) as
\begin{align}
	&\frac{\partial }{\partial r}\left[ nu^r(r^2+a^2)\sin \theta \right]+\frac{\partial}{\partial \theta}\left[ nu^{\theta}(r^2+a^2)\sin\theta \right]=0,
	\label{eq:continuity equation for axis symmetrical steady-state}\\
	&\frac{\partial u^t}{\partial r}u^r+\frac{\partial u^t}{\partial \theta}u^{\theta}=0,
	\label{eq:Euler equations for axis symmetrical steady-state t}\\
	&\frac{\partial u^r}{\partial r}u^r+\frac{\partial u^r}{\partial \theta}u^{\theta}-r(u^{\theta})^2-r\sin^2\theta (u^{\varphi})^2=0,
	\label{eq:Euler equations for axis symmetrical steady-state r}\\
	&(r^2+a^2)\frac{\partial u^{\theta}}{\partial r}u^r+(r^2+a^2)\frac{\partial u^{\theta}}{\partial \theta}u^{\theta}\notag \\
	&\qquad+2ru^{\theta}u^r-(r^2+a^2)\sin\theta\cos\theta(u^{\varphi})^2=0,
	\label{eq:Euler equations for axis symmetrical steady-state theta}\\
	&(r^2+a^2)\sin^2\theta\frac{\partial u^{\varphi}}{\partial r}u^r+(r^2+a^2)\sin^2\theta\frac{\partial u^{\varphi}}{\partial \theta}u^{\theta}\notag \\
	&+2r\sin^2\theta u^{\varphi}u^r+2(r^2+a^2)\sin\theta\cos\theta u^{\varphi}u^{\theta}=0.
	\label{eq:Euler equations for axis symmetrical steady-state phi}
\end{align}

Here we assume that dust rotates so slowly that the solution of $u^\mu$ and $n$ are given by perturbations about the spherically symmetric solution obtained in Ref.\cite{spherical steady-state solutions} as follows.
\begin{align}
	&u^t=u^t_0+u^t_1(r,\,\theta),
	\label{eq:perturbation of solution u^t}\\
	&u^r=u^r_0+u^r_1(r,\,\theta),
	\label{eq:perturbation of solution u^r}\\
	&u^{\theta}=u^{\theta}_1(r,\,\theta),
	\label{eq:perturbation of solution u^theta}\\
	&u^{\varphi}=u^{\varphi}_1(r,\,\theta),
	\label{eq:perturbation of solution u^phi}\\
	&n=n_{a}\frac{a^2}{r^2+a^2}+n_1(r,\,\theta),
	\label{eq:perturbation of solution n}
\end{align}
where
$u^t_0,\,u^r_0$ and $n_a$ are constants of the unperturbed (i.e., spherically symmetric) solutions, and 
$u^t_1,\,u^r_1,~u^\theta_1,~u^\varphi_1$ and $n$ are perturbed quantities caused by dust rotation.

We substitute these expressions into Eqs.(\ref{eq:continuity equation for axis symmetrical steady-state}) - (\ref{eq:Euler equations for axis symmetrical steady-state phi}) and solve them up to the first order of the perturbed quantities. Then we obtain the solution of  $u^t_1,\,u^r_1,\,u^{\theta}_1,\,u^{\varphi}_1$ and $n_1$ as 
\begin{align}
	&u^t_1=\frac{u^r_0}{u^t_0}f^r(\theta),
	\label{eq:solution of u^t_1}\\
	&u^r_1=f^r(\theta),
	\label{eq:solution of u^r_1}\\
	&u^{\theta}_1=\frac{a^2}{r^2+a^2}f^{\theta}(\theta),
	\label{eq:solution of u^theta_1}\\
	&u^{\varphi}_1=\frac{a^2}{r^2+a^2}f^{\varphi}(\theta),
	\label{eq:solution of u^phi_1}\\
	&n_1=\frac{n_a}{u^r_0}\frac{a^2}{r^2+a^2}\left\{ a\cdot\tan^{-1}\left( \frac{r}{a} \right)+C(\theta) \right\}N(\theta),
	\label{eq:solution of n_1}
\end{align}
where $f^r,\,f^{\theta},\,f^{\varphi}$ and $C$ are arbitrary functions of $\theta$, and $N$ is defined as
\begin{align}
	&N(\theta)\equiv-\frac{1}{\sin\theta}\frac{\partial}{\partial \theta}\left[ f^{\theta}\sin\theta \right].
	\label{eq:N(theta)}
\end{align}

From the regularity condition on the poles $(\theta=0,\pi)$, $N(\theta)$ must satisfy
\begin{align}
	&\left.\frac{dN}{d\theta}\right|_{\theta=0}=\left.\frac{dN}{d\theta}\right|_{\theta=\phi}=0.
	\label{eq:condition expressions for N}
\end{align}
As a solution of $f(\theta)$ which satisfies (\ref{eq:condition expressions for N}), we adopt
\begin{align}
	&f^{\theta}(\theta)=f^{\theta}_0\sin(k\theta),\quad \{k\in \mathbb{Z}^+\},
	\label{eq:supposed solution of f^theta}
\end{align}
where $f^{\theta}_0$ is a dimensionless constant.
Then, we can rewrite Eq.(\ref{eq:N(theta)}) as
\begin{align}
	&N(\theta)=-f^{\theta}_0\left[ k\cos(k\theta)+\sin(k\theta)\cot\theta \right].
	\label{eq:supposed solution of N(theta)}
\end{align}
Here we suppose that dust is distributed reflection-symmetrically with respect to the $\theta=\pi/2$ plane, that is, $k$ is even.

\begin{figure*}
	\includegraphics[width=.75\hsize]{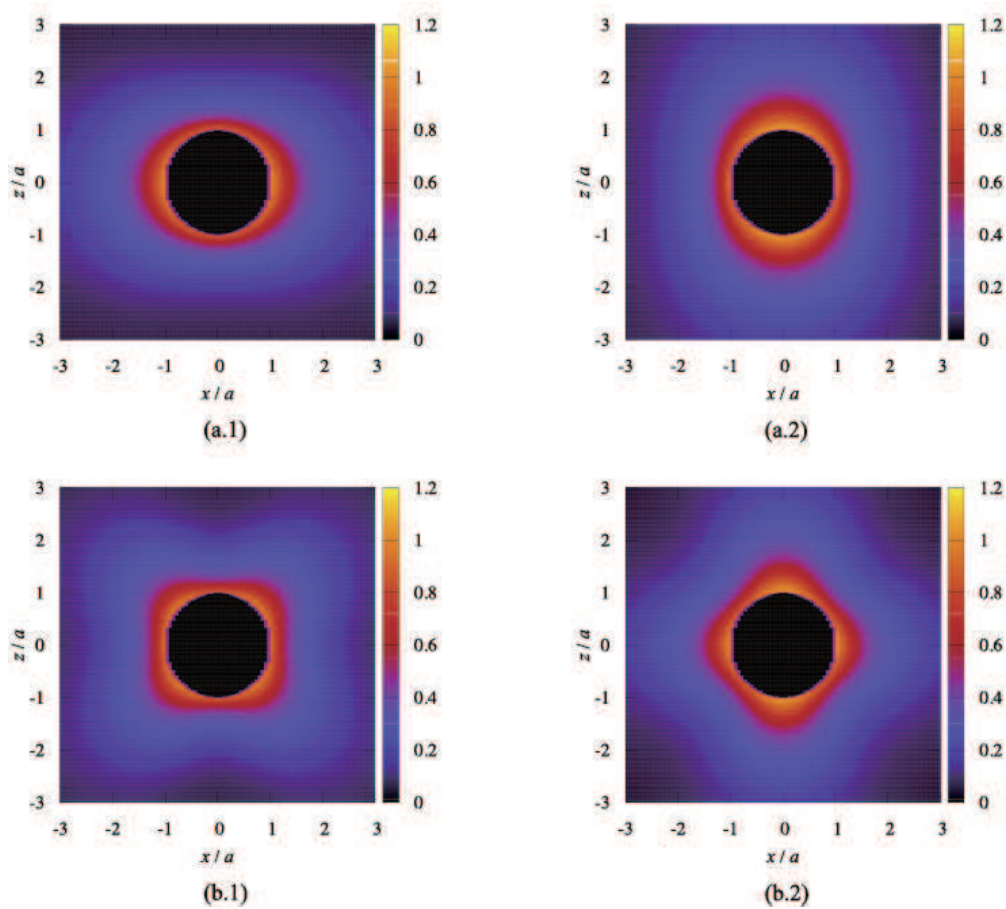}
	\caption{	\label{fig:solutions of dust density}
Contour map of the dust density $n(r,\,\theta)/n_a$.
The rectangle coordinates $(x,\,z)$ are defined as (\ref{xz}) and hence 
the wormhole throat corresponds to the circle $x^2+z^2=a^2$.
No real spacetime exists in the black region of $x^2+z^2<a^2$.
We set $u^r_0=-0.1$ and $C(\theta)=0$, and illustrate four cases,
$(f^{\theta}_0,\,k)=(-0.02,\,2),\,(0.02,\,2),\,(-0.02,\,4)$ and $(0.02,\,4)$, in (a.1), (a.2), (b.1), and (b2), respectively.
}
\end{figure*}

Figure \ref{fig:solutions of dust density} is the contour map of the dust density $n(r,\,\theta)/n_a$.
The rectangle coordinates $(x,\,z)$ are defined as 
\begin{equation}\label{xz}
x=\sqrt{r^2+a^2}\sin\theta,~~~z=\sqrt{r^2+a^2}\cos\theta.
\end{equation}
This means that the wormhole throat corresponds to the circle $x^2+z^2=a^2$.
No real spacetime exists in the black region of $x^2+z^2<a^2$.
We set $u^r_0=-0.1$ and $C(\theta)=0$, and illustrate four cases,
$(f^{\theta}_0,\,k)=(-0.02,\,2),\,(0.02,\,2),\,(-0.02,\,4)$ and $(0.02,\,4)$.
We find the following characteristics, 
which are caused by the effect of rotation.
\begin{itemize}
\item While the density peak is located at the throat $(r=a)$ in the case of spherically symmetry, 
it deviates slightly to the outer side of the throat in the present case.
\item The high density region is not spherical but distorted; the shape varies depending on the values of $(f^{\theta}_0,\,k)$.
\end{itemize}

\section{wormhole shadows}

We investigate the optical images of dust surrounding the wormhole, using the axially symmetric solutions obtained in Sec. IV.

\subsection{Apparent position of optical sources}

\begin{figure}
\begin{center}
          \includegraphics[width=.8\hsize]{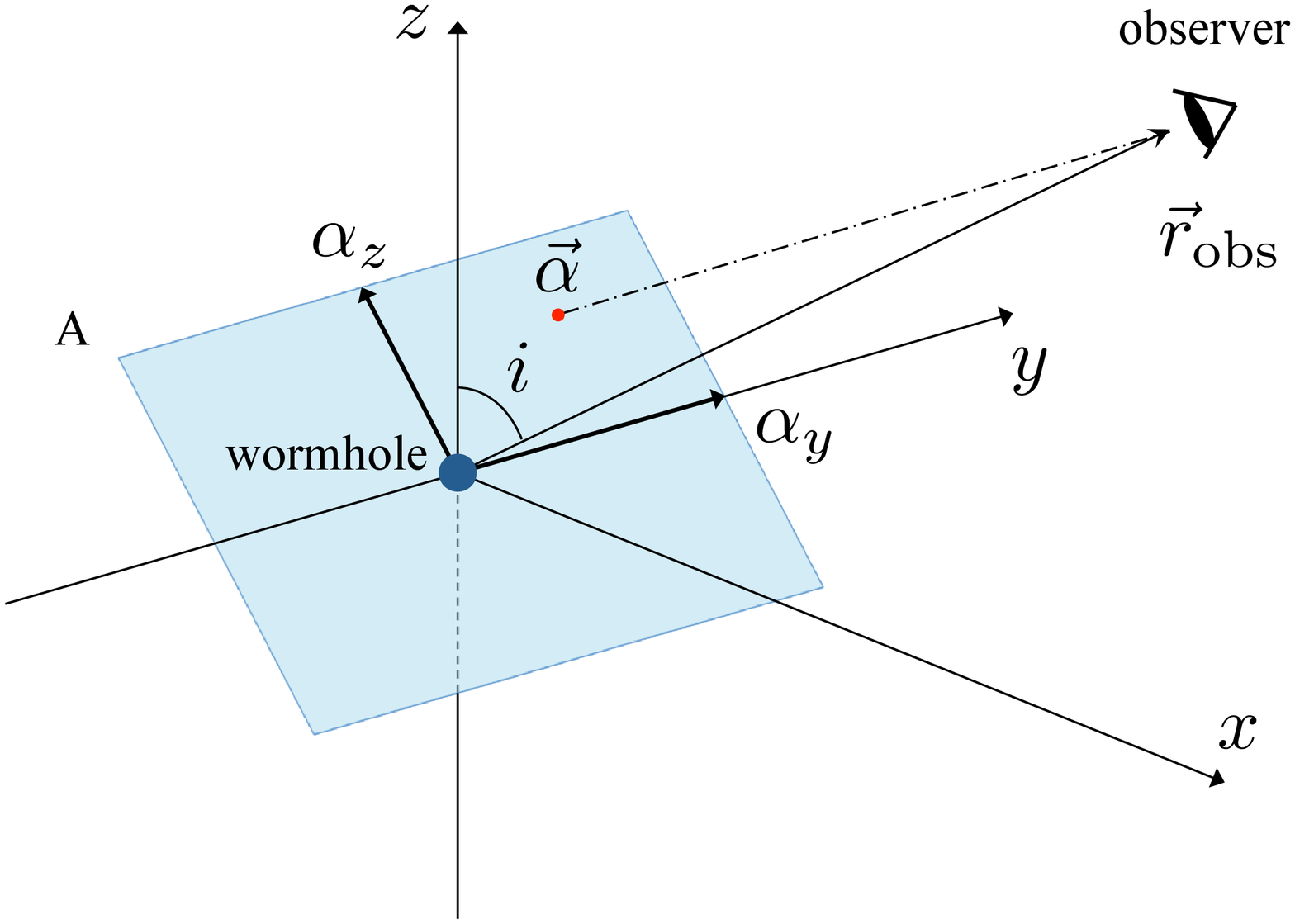}\\\vspace{.02\vsize}
          (a)\\ \vspace{.02\vsize}
          \includegraphics[width=.7\hsize]{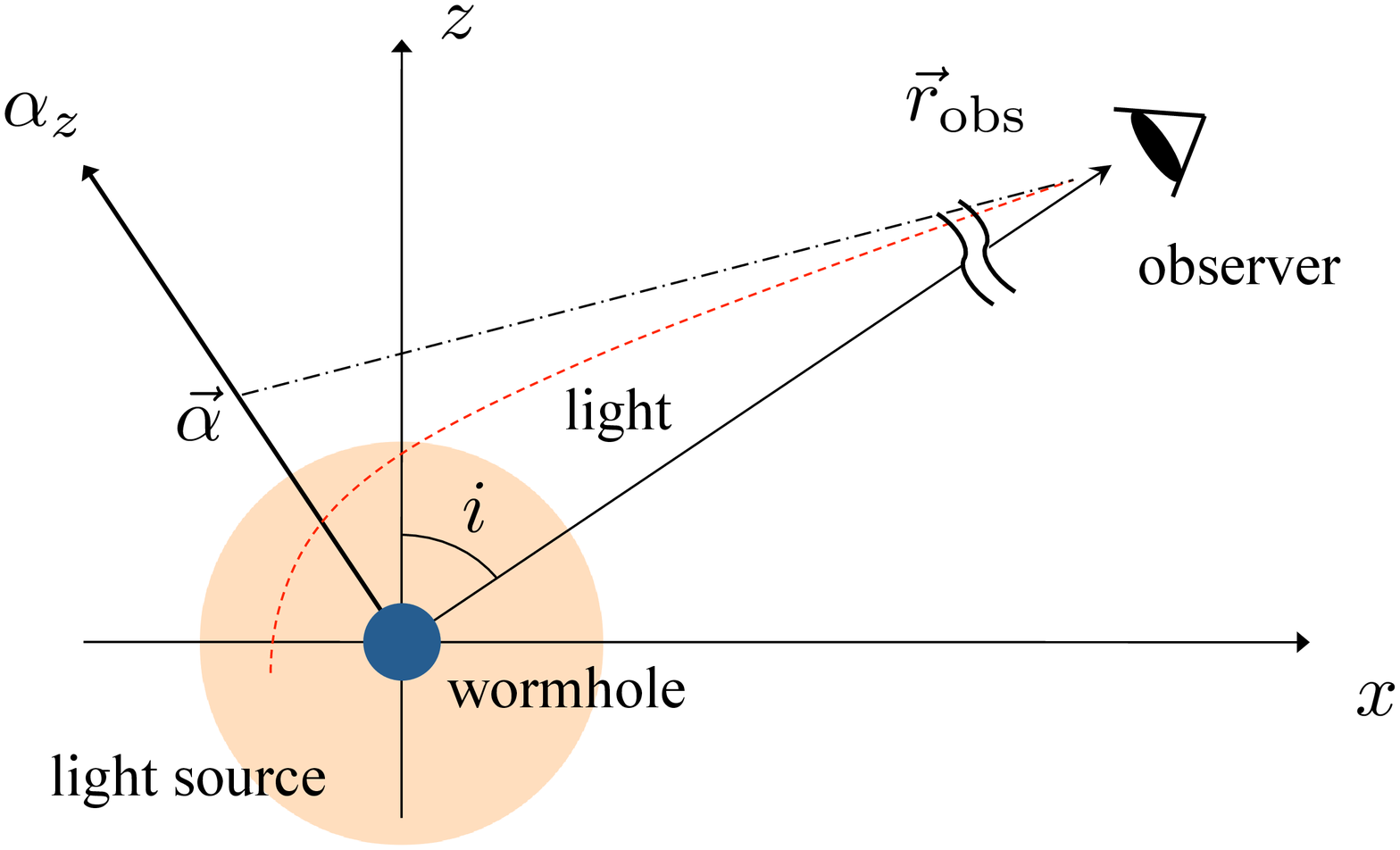}\\\vspace{.02\vsize}
          (b)\\
 \end{center}
    \caption{
{Positional relationship among the wormhole, the apparent position of a light source, and the observer.
(a) Plane A is defined as the plane which is normal to $\vec{r}_{\text{obs}}$ and contains the origin.
We denote the intersection of Plane A with the tangent to the ray at the observer by $\vec\alpha$, which corresponds the apparent position of a light source.
(b) is the same diagram as (a) but the standpoint is on the $y$-axis.
We accord the rotational axis of dust to $z$-axis and denote the angle between it and $\vec{r}_{\text{obs}}$ as $i$.
We put the observer on the $x$-$z$ plane.
}}
    \label{fig:apparent position of light source}
\end{figure}

Gravity refracts light rays in general relativity.
This phenomenon is known as gravitational lensing.
We may see a light source as if it was another position.
To obtain optical images, we have to not only solve the null geodesic equations, but also give the relation between the null vector at the observer and the apparent potion of the source.

We put an Ellis wormhole at the origin, an observer at $\vec{r}_{\text{obs}}$, as shown in Fig. \ref{fig:apparent position of light source}.
Plane A is defined as the plane which is normal to $\vec{r}_{\text{obs}}$ and contains the origin.
We denote the intersection of Plane A with the tangent to the ray at the observer by $\vec\alpha$, which corresponds the apparent position of a light source, as shown in (a).
Figure \ref{fig:apparent position of light source} (b) is the same diagram as (a) but the standpoint is on the $y$-axis.
We accord the rotational axis of dust to $z$-axis and denote the angle between it and $\vec{r}_{\text{obs}}$ as $i$.
We put the observer on the $x$-$z$ plane.

Then the two vectors $\vec{r}_{\text{obs}}$ and $\vec{\alpha}$ are expressed as
\begin{align}
	&\vec{r}_{\text{obs}}=(r_{\text{obs}}\sin i,~0,~r_{\text{obs}}\cos i),
	\label{eq:position of observer}\\
	&\vec{\alpha}=(-\alpha_z\cos i,~\alpha_y,~\alpha_z\sin i),
	\label{eq:apparent position of light source}
\end{align}
where $r_{\text{obs}}\equiv |\vec{r}_{\text{obs}}|$.
Thus, the equations of the line which connects the two points are written as
\begin{align}
	\frac{x-{r}_{\text{obs}}\sin i}{{r}_{\text{obs}} \sin i+\alpha_z\cos i}=-\frac{y}{\alpha_y}=\frac{z-{r}_{\text{obs}}\cos i}{{r}_{\text{obs}}\cos i-\alpha_z\sin i}.
	\label{eq:apparent line of light}
\end{align}
Differentiating the equations and taking the limit $(r\rightarrow \vec{r}_{\text{obs}},\,\theta\rightarrow i,\,\varphi\rightarrow 0)$, we obtain
\begin{align}
	&\left( \sin i+{r}_{\text{obs}}\cos i\,\frac{d\theta}{dr} \right)\alpha_y({r}_{\text{obs}}\cos i-\alpha_z\sin i)\notag \\
	&=-{r}_{\text{obs}}\sin i\,\frac{d\varphi}{dr}({r}_{\text{obs}}\sin i+\alpha_z\cos i)({r}_{\text{obs}}\sin i+\alpha_z\sin i)\notag \\
	&=\left( \cos i-{r}_{\text{obs}}\sin i\,\frac{d\theta}{dr} \right)\alpha_y({r}_{\text{obs}}\sin i+\alpha_z\cos i).
	\label{eq:def. and limit of apparent line of light}
\end{align}
We solve these equations for $\alpha_y$ and $\alpha_z$:
\begin{align}
	&\alpha_y=-{{r}_{\text{obs}}}^2\sin i\,\frac{d\varphi}{dr}=-{{r}_{\text{obs}}}^2\sin i\,\frac{k^{\varphi}(\vec{r}_{\text{obs}})}{k^r(\vec{r}_{\text{obs}})},
	\label{eq:solution for apparent position of alpha y}\\
	&\alpha_z={{r}_{\text{obs}}}^2\frac{d\theta}{dr}={{r}_{\text{obs}}}^2\frac{k^{\theta}(\vec{r}_{\text{obs}})}{k^r(\vec{r}_{\text{obs}})}.
	\label{eq:solution for apparent position of alpha z}
\end{align}
We thus obtain the relation between the apparent position of a light source and the null vector at the observer.

\subsection{Radiation intensity}
To calculate the observed intensity of radiation emitted from optically thin gas, we solve the general relativistic radiative transfer equation, which is generally expressed as  \cite{FOUNDATIONS OF RADIATION HYDRODYNAMICS},
\begin{align}
	\frac{d\, \mathfrak{J}}{d\lambda}=\frac{\eta(\nu)}{\nu^2}-\nu\chi(\nu)\mathfrak{J},\qquad \mathfrak{J}\equiv\frac{I(\nu)}{\nu^3}
	\label{eq:radiative transport equation}
\end{align}
where $\nu$ is the photon frequency, $I(\nu)$ is the specific intensity, $\mathfrak{J}$ is the invariant intensity, $\eta(\nu)$ is the emission coefficient and $\chi(\nu)$ is the absorption coefficient.
Because Eq.(\ref{eq:radiative transport equation}) is the differential equation along null geodesics, we should solve the null geodesic equations simultaneously.

Here we make the following assumptions, for simplicity:
\begin{itemize}
\item The dust does not absorb radiation, i.e., $\chi(\nu)=0$.
\item $\eta(\nu)$ is proportional to the dust density which is measured along the null geodesics, i.e., $\eta(\nu) d\lambda \propto\rho u_\mu dx^{\mu}$.
\end{itemize}
Introducing a positive factor $H(\nu)$, which is proportional to the spectrum of the dust sources, we express $\eta(\nu)$ as
\begin{align}
	\eta(\nu)d\lambda =-H(\nu)\rho u_{\mu}dx^{\mu}.
	\label{eq:radiation term}
\end{align}
With these assumptions we can integrate (\ref{eq:radiative transport equation}) as
\begin{align}
	\mathfrak{J}=-\int\frac{H(\nu)}{\nu^2}\rho u_{\mu}dx^{\mu}.
	\label{eq:integration of transfer equation}
\end{align}
The integration in Eq.(\ref{eq:integration of transfer equation}) should be performed alongside the null geodesics.
The frequency measured by observers comoving with dust particles is given by
\begin{equation}
	\nu=-u_\mu k^\mu.
	\label{eq:frequency at each point}
\end{equation}

Generally we should fix the spectrum of the dust sources, i.e., $H(\nu)$.
Here, for simplicity, we assume a flat spectrum, $H(\nu)=\text{const}$.

\subsection{Numerical analysis}

We compute the intensity distribution as follows:
\def\theenumi{(\roman{enumi})}
\begin{enumerate}
	\item Put the observer at $r_{\text{obs}}=300a$ and $i=\pi/2$.
	\item For given $\vec{\alpha}$, we solve Eqs.(\ref{eq:solution for apparent position of alpha y}) and (\ref{eq:solution for apparent position of alpha z}) to obtain $k^{\varphi}(\vec{r}_{\text{obs}})$  and $k^{\theta}(\vec{r}_{\text{obs}})$ with fixing $k^r(\vec{r}_{\text{obs}})=1.0$.
Then we solve the null geodesic equations from the observer and obtain the frequencies $\nu$ at each points by Eq.(\ref{eq:frequency at each point}).
We can choose a value of the initial (observed) frequency $\nu_{\text{obs}}$ arbitrarily because the ratio of $\nu_{\text{obs}}$ to the emitted frequency $\nu_{\text{emit}}$ does not depend on $\nu_{\text{obs}}$.
	\item With the values of $\nu$ at each point, which is determined by the null geodesic equations, we integrate Eq.(\ref{eq:integration of transfer equation}) to obtain the intensity $I$. We adopt the fourth-order Runge-Kutta method for all integrations.
	\item We continue the integrations until $r_{\text{obs}}=300a$ again, where the gas density is sufficiently small.
	\item Iterate (ii) $\sim$ (iv) by changing the value of $\vec{\alpha}$.
\end{enumerate}
We adopt the dust solutions (\ref{eq:solution of u^t_1}) - (\ref{eq:solution of n_1}) with the parameters, $u^r_0=-0.1$, $f^r=0$, $f^{\varphi}=0.02$ and $C(\theta)=0$.
        We calculate for four cases, $(f^{\theta}_0,\,k)=(-0.02,\,2),\,(0.02,\,2),\,(-0.02,\,4)$ and $(0.02,\,4)$.

\begin{figure*}
	\includegraphics[width=.75\hsize]{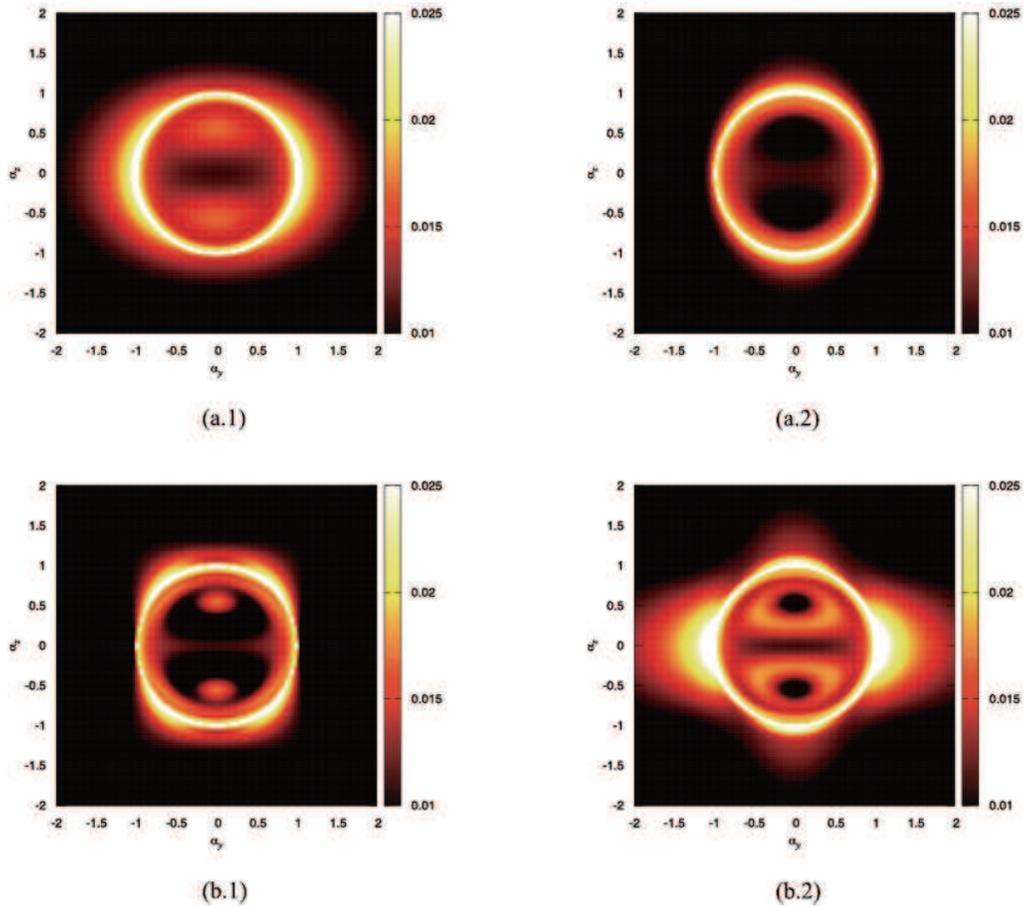}
	\caption{ \label{fig:numerical solutions}
Optical images of the wormhole surrounded by rotating dust for several cases of $(f^{\theta}_0,\,k)$: (a.1), (a.2), (b.1) and (b.2) show the cases of $(-0.02,\,2),\,(0.02,\,2),\,(-0.02,\,4)$ and $(0.02,\,4)$, respectively. 
A bright ring looks distorted like the high density region of dust.
Additionally, there appears a weakly luminous pattern differently depending on the value of $(f^{\theta}_0,\,k)$.
	}
\end{figure*}
\begin{figure*}
	\includegraphics[width=.75\hsize]{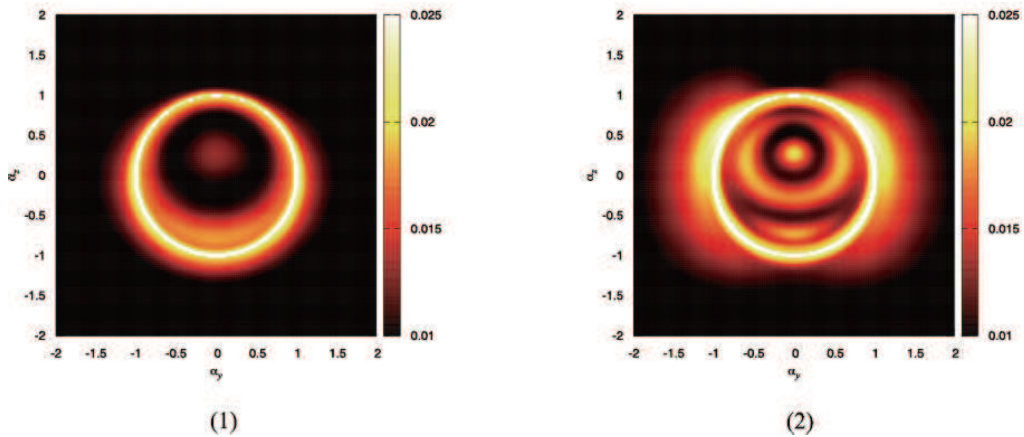}
	\caption{\label{fig:numerical solutions 2}
Optical images on the case of $i=\pi/4$: (1) is $(f^{\theta}_0,\,k)=(-0.02,\,2)$, (2) is $(f^{\theta}_0,\,k)=(-0.02,\,4)$.
Other parameters are same values as the case of Fig.\ref{fig:numerical solutions}.
The patterns are more complex than it on the case of Fig.\ref{fig:numerical solutions}.
	}
\end{figure*}

We show the results of our numerical calculations in Fig. \ref{fig:numerical solutions}.
We find the following phenomena.
\begin{itemize}
\item A bright ring appears on each case.
\item The bright ring looks distorted due to rotation.
\item There appear weakly luminous complex patterns, which vary depending on the values of $(f^{\theta}_0,\,k)$.
\end{itemize}
The first phenomenon is due to the existence of unstable circular orbits of photons.
The second phenomenon occurs because the equatorial region is more dense, as discussed in Sec. IV.C.
The third phenomenon is due to the existence of the other side region of the wormhole.

Figure \ref{fig:numerical solutions 2} shows the different case from the above computations.
Here, we compute on the case of $i=\pi/4$, and other parameters are same as Figs. \ref{fig:numerical solutions}(a.1) and (b.1): (1) is $(f^{\theta}_0,\,k)=(-0.02,\,2)$ and (2) is $(f^{\theta}_0,\,k)=(-0.02,\,4)$.
The patterns are more complex than it on the case of Fig.\ref{fig:numerical solutions}.
%%%%%%%%%%%%%%%%%%%%%%%%%%%%%%%%
%
%
%
%
%%%%%%%%%%%%%%%%%%%%%%%%%%%%%%%%
\section{Concluding remarks}
This work is an extension of our previous work which studied wormhole shadows in nonrotating dust.
In the case of spherically symmetric dust, we had found two types of solutions: $u^r=0$ and $\rho=$ arbitrary function of $r$ (Type I) and $u^r\ne0,~\rho\propto (r^2+a^2)^{-1}$ (Type II).
First, we derived steady-state solutions of rotating dust and more general medium surrounding the wormhole by solving general relativistic Euler equations.
We consider the perturbation from the type II of nonrotating solution.
Our solutions for rotating dust density (\ref{eq:perturbation of solution n}) with (\ref{eq:solution of n_1}) has a peak at a little outside of the throat $(r>a)$.
This is the clearly different point from the nonrotating dust.

Next, solving radiation transfer equations along null geodesics, we investigated the optical images of the wormhole surrounded by dust for the above steady-state rotating solutions.
A bright ring appears in the image, just as in Schwarzschild spacetime, because the wormhole spacetime possesses unstable circular orbits of photons.
The bright ring looks distorted because the equatorial region is more dense.
Additionally, there appears a weakly luminous pattern differently depnding on on the parameters of $(f^{\theta}_0,\,k)$.
This pattern is due to the existence of the other side region of the wormhole.

Because our method is general, it is applicable to other astronomical bodies (e.g. gravastars \cite{gravastar}) for researches of these optical images.
Shodows of wormholes and other compact objects could be detected with high-resolution VLBI observations in the near future.
%%%%%%%%%%%%%%%%%%%%%%%%%%%%%%%%
%
%
%
%
%
%%%%%%%%%%%%%%%%%%%%%%%%%%%%%%%%
\acknowledgements
We thank F. Abe, K. Fujisawa, T. Harada, H. Saida and K. Shiraishi for useful discussions.

%%%%%%%%%%%%%%%%%%%%%%%%%%%%%%%%
%
%
%
%
%
%%%%%%%%%%%%%%%%%%%%%%%%%%%%%%%%%%%%%%%%%%%%%%%%

%%%%%%%%%%%%%%%%%%%%%%%%%%%%%%%%%%%%%%%%%%%%%%%%
\end{document}